\documentclass{article}

\usepackage{amsmath,amssymb,amsfonts}   
\usepackage{graphicx,bm}

\newcommand{\calU}{{\mathcal U}}
\newcommand{\R}{{\mathbb R}}

\newcommand{\X}{\mathbf{X}}
\renewcommand{\P}{\mathbb{P}}
\newcommand{\x}{\mathbf{x}}
\newcommand{\y}{\mathbf{y}}
\newcommand{\e}{{\mathrm e}}
\newcommand{\E}{{\mathbb E}}

\newcommand{\n}{\mathbf n}
\newcommand{\calT}{{\mathcal T}}
\renewcommand{\P}{\mathbb P}
\newcommand{\p}{\widetilde{p}}
\newcommand{\q}{{q}}

\begin{document}

\title{\bf Asymptotic analysis of extended two-dimensional narrow capture problems}

\author{ \em
P. C. Bressloff, \\ Department of Mathematics, 
University of Utah \\155 South 1400 East, Salt Lake City, UT 84112}

 \maketitle


\begin{abstract} In this paper we extend our recent work on two-dimensional (2D) diffusive search-and-capture processes with multiple small targets (narrow capture problems) by considering an asymptotic expansion of the Laplace transformed probability flux into each target. The latter determines the distribution of arrival or capture times into an individual target, conditioned on the set of events that result in capture by that target. A characteristic feature of strongly localized perturbations in 2D is that matched asymptotics generates a series expansion in $\nu=-1/\ln \epsilon$ rather than $\epsilon$, $0<\epsilon\ll 1$, where $\epsilon$ specifies the size of each target relative to the size of the search domain. Moreover, it is possible to sum over all logarithmic terms non-perturbatively. We exploit this fact to show how a Taylor expansion in the Laplace variable $s$ for fixed $\nu$ provides an efficient method for obtaining corresponding asymptotic expansions of the splitting probabilities and moments of the conditional FPT densities. We then use our asymptotic analysis to derive new results for two major extensions of the classical narrow capture problem: optimal search strategies under stochastic resetting, and the accumulation of target resources under multiple rounds of search-and-capture.

\end{abstract}


\section{Introduction}

There are numerous biological, social, and physical processes, where the
arrival of a single particle (searcher) at a target site
can initiate one or more downstream events. Examples include the provision of food or shelter to a foraging animal \cite{Bell91,Bartumeus09,Viswanathan11}, the initiation of a biochemical reaction within a cell \cite{Loverdo08,Benichou10}, the entry of a virus into the cell nucleus \cite{Damm06,Lagache08,Lawley15}, the delivery of proteins to a subcellular compartment by molecular motor complexes \cite{Bressloff13}, and the initiation of an immune response due to a T cell finding its antigen \cite{Coombs15}. 
These so-called search-and-capture events are often stochastic in nature, and crucial information regarding the efficacy of the search process can be obtained from the distribution of arrival or capture times, which is known as the first-passage-time (FPT)  density. Moments of the FPT density typically satisfy corresponding boundary value problems (BVPs), which can be derived from the backward evolution equation for the probability density of particle position. In many cases, there are multiple targets within
the interior or on the boundary of the search domain, which requires determining the splitting probability of being captured by a specific target. Since this probability is less than unity, it follows that the corresponding FPT density has infinite moments, unless it is conditioned on the set of events that find the target. The classical narrow capture (or escape) problem concerns diffusive search processes where the targets are much smaller than the size of the search domain. This then allows matched asymptotic expansions and Green's functions to be used to solve the BVPs for the splitting probabilities and moments of the conditional FPT density \cite{Ward93,Schuss07,Bressloff08,Coombs09,Cheviakov11,Chevalier11,Holcman14,Coombs15,Ward15,Bressloff15,Lindsay15,Lindsay16,Lindsay17}. 

As we have recently highlighted \cite{Bressloff20}, there are growing number of search problems that require knowing the full conditional FPT densities, or at least their Laplace transforms, rather than simply the first and second moments, say. For example, one way to increase the efficacy of a search process is to include a resetting protocol, whereby the position of the searcher is reset to a fixed location $\x_r$ at a random sequence of times, which is typically (but not necessarily) generated by a Poisson process. In many cases there exists an optimal resetting rate for minimizing the mean first passage time (MFPT) to reach a target, see the review \cite{Evans20} and references therein. Assuming that once the particle has returned to $\x_r$ it loses all memory of previous search phases, that is, the stochastic process satisfies the strong Markov property, one can condition on whether or not the particle resets at least once, even though a reset event occurs at random times. Renewal theory can then be used to express statistical quantities with resetting in terms of statistical quantities without resetting such as the conditional FPT densities. This has been carried out both for one or two targets \cite{Reuveni16,Pal17,Belan18,Pal19,Pal20,Bodrova20} and multiple targets \cite{Bressloff20,Bressloff20m}. A second example occurs in cases such as the intracellular transport and delivery of vesicular cargo to subcellular compartments, where one has to keep track of the accumulation of resources within the targets following multiple rounds of search-and-capture events and degradation. Queuing theory can be used to express moments of the steady-state distribution of resources across the target population in terms of the conditional FPT densities of a single search-and-capture event \cite{Bressloff19,Bressloff20a,Bressloff20}.

In this paper we extend our recent work on two-dimensional (2D) diffusive search-and-capture processes with multiple small targets by considering asymptotic expansions of the conditional FPT densities. Following along similar lines to \cite{Lindsay16}, we work directly with the Laplace transform of the forward diffusion equation, which is solved by constructing an inner or local solution
valid in an $O(\epsilon)$ neighborhood of each target, and then matching to an outer or global solution that is valid away from each neighborhood. Here $\epsilon$ is a small dimensionless parameter that characterizes the size of each target relative to the size of the search domain. It is well known that since the 2D Green's function of the diffusion equation has a logarithmic singularity, it is more natural to develop an asymptotic expansion in $\nu= -1/\ln \epsilon$ rather than $\epsilon$ itself. Indeed, it is possible to sum over the logarithmic terms non-perturbatively \cite{Ward93}. This is equivalent to calculating the asymptotic solution for all terms of $O(\nu^k)$ for any $k$. Having solved the diffusion equation in Laplace space, we determine the corresponding Laplace transforms of the probability fluxes into each target and hence the conditional FPT densities. We then show how a Taylor expansion of the FPT densities in the Laplace variable $s$ for fixed $\nu$ provides an efficient method for obtaining corresponding asymptotic expansions of the splitting probabilities and FPT moments, without having to solve separate boundary value problems. Finally, our analysis is used to obtain new results concerning optimal search strategies under stochastic resetting, and the accumulation of target resources under multiple rounds of search-and-capture.
 First, we determine the effects of stochastic resetting on the unconditional MFPT, extending previous results in the small-$r$ regime. We proceed by exploiting the exponential-like asymptotic decay of the Green's function for the modified Helmholtz equation, in order to construct boundary-free approximations of statistical quantities that hold for intermediate and large values of the resetting rate $r$. This allows us to identify target configurations where the MFPT is minimized at an optimal resetting rate. Second, we investigate the size of fluctuations in the steady-state distribution of target resources under multiple rounds of search-and-capture, by expressing the Fano factor in terms of the splitting probabilities and unconditional MFPTs for a search process with resetting at the degradation rate $\gamma$.
The paper is structured as follows. The classical narrow capture problem is presented in section 2, and the matched asymptotics is carried out in section 3. Applications to narrow capture with stochastic resetting and to target resource accumulation are then developed in sections 4 and 5, respectively.

\section{Classical narrow capture problem}

\begin{figure}[b!]
\begin{center} 
\includegraphics[width=7cm]{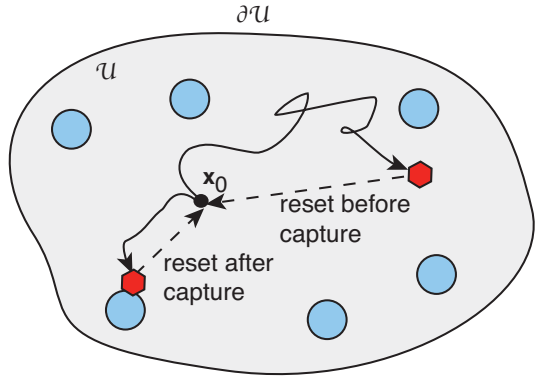} 
\caption{Diffusion of a particle in a bounded domain $\calU$ with $N$ small targets $\calU_j$, $j=1,\ldots,N$. If the particle hits a boundary $\partial\calU_j$ then it is absorbed. Also shown are two extensions of the classical narrow capture problem that are considered in sections 4 and 5. {\bf Reset before capture}: at any time the particle can return to its initial position $\x_0$ at a Poisson rate $r$ and immediately restart the search process (stochastic resetting protocol). {\bf Reset after capture}: each time the particle is captured by a target, it delivers some resource to the target, after which it returns to $\x_0$ and initiates another round of search-and-capture (target resource accumulation).}
\label{fig1}
\end{center}
\end{figure}

Consider a two-dimensional bounded domain $\calU\subset \R^2$ that contains a set of $N$ small interior targets $\calU_k$, $k=1,\ldots,N$, with $\bigcup_{j=1}^N \calU_k=\calU_a\subset \calU$, see Fig. \ref{fig1}. In the absence of any resetting events, we have a classical narrow escape problem.
Let $p(\x,t|\x_0)$ be the probability density that at time $t$ a particle is at $\X(t)=\x$, having started at position $\x_0$. Then
\begin{subequations} 
\label{master}
\begin{align}
	\frac{\partial p(\x,t|\x_0)}{\partial t} &= D\nabla^2 p(\x,t|\x_0), \ \x\in \calU\backslash \calU_a,\quad \nabla p \cdot \n=0, \ \x\in \partial \calU,\\
	p(\x,t|\x_0) &=0,\  \x\in \partial\calU_a,
	\end{align}
	\end{subequations} 
together with the initial condition $p(\x,t|\x_0)=\delta(\x-\x_0)$.
Each target is assumed to have a size $|\calU_j|\sim \epsilon^2 |\calU|$ with $\calU_j\rightarrow \x_j\in \calU$ uniformly as $\epsilon \rightarrow 0$, $j=1,\ldots,N$. The targets are also taken to be well separated in the sense that $|\x_i-\x_j|=O(1)$, $j\neq i$, and $\mbox{dist}(x_j,\partial \calU)=O(1)$.
For the sake of illustration, we take each target to be a circle of radius $\epsilon \ell$ and fix length scales by setting $\ell =1$. Thus $\calU_i=\{\x \in \calU, \ |\x-\x_i|\leq \epsilon\}$.

The probability flux into the $k$-th target at time $t$ is 
\begin{align}
\label{J}
	J_k(\x_0,t)&=-D \int_{\partial \calU_k} \nabla p(\x,t|\x_0)\cdot \n d\sigma,\ k = 1,\ldots,N,
	\end{align}
	where $\n$ is the inward normal to $\calU_k$.
Hence, the probability that the particle is captured by the $k$-th target after time $t$ is
\begin{equation}
\label{Pi}
\Pi_k(\x_0,t)=\int_t^{\infty}J_k(\x_0,t')dt' ,
\end{equation}
and the corresponding splitting probability is
\begin{equation}
\label{split}
\pi_k(\x_0)=\Pi_k(\x_0,0)=\int_0^{\infty}J_k(\x_0,t')dt' =\widetilde{J}_k(\x_0,0).
\end{equation}
It follows that the Laplace transform of $\Pi(\x_0,t)$ is given by
\begin{align}
\label{Pik}
s\widetilde{\Pi}_k(\x_0,s)-\pi_k=-\widetilde{J}_k(\x_0,s)=D\int_{\partial \calU_k} \nabla \widetilde{p}(\x,s|\x_0)\cdot \n d\sigma.
\end{align}
Next we introduce the survival probability that the particle hasn't been absorbed by a target in the time interval $[0,t]$, having started at $\x_0$:
\begin{equation}
\label{Q1}
Q(\x_0,t)=\int_{\calU\backslash \calU_a}p(\x,t|\x_0)d\x.
\end{equation}
Differentiating both sides of this equation with respect to $t$ and using Eqs. (\ref{master}) implies that
\begin{align}
\frac{\partial Q(\x_0,t)}{\partial t}&=D\int_{\calU\backslash \calU_a}\nabla\cdot \nabla p(\x,t|\x_0)d\x=D\sum_{k=1}^N \int_{ \partial \calU_k}\nabla p(\x,t|\x_0)\cdot \n d\sigma\nonumber \\
& =-\sum_{k=1}^NJ_k(\x_0,t).
\label{Q2}
\end{align}
Laplace transforming equation (\ref{Q2}) gives
\begin{equation}
\label{QL}
s\widetilde{Q}(\x_0,s)-1=- \sum_{k= 1}^N \widetilde{J}_k(\x_0,s).
\end{equation}
We have used the initial condition $Q(\x_0,0)=1$. It immediately follows from equation (\ref{split}) that
\begin{equation}
\sum_{k=1}^N\pi_k(\x_0)=1.
\end{equation}

Since the probability of the particle being captured by the $k$-th target is typically less than unity ($\pi_k<1$), it follows that the moments of the corresponding FPT density are infinite unless we condition on the given event. The MFPT $\calT_k$ to be captured by the $k$-th target is given by
\begin{equation}
\calT_k(\x_0)=\inf\{t>0; \X(t)\in \partial\calU_k|\X(0)=\x_0\},
\end{equation}
with $\calT_k=\infty$ if the particle is captured by another target. Introducing the set of events $\Omega_k=\{\calT_k<\infty\}$, we can then define the conditional FPT density according to
\begin{align*}
f_k(\x_0,t)dt&=\P[t<\calT_k<t+dt|\calT_k<\infty,\X(0)=\x_0]\\
&=\P[t<\calT_k<t+dt|\X(0)=\x_0]/\P[\Omega_k] \nonumber \\
&=(\Pi_k(\x_0,t]-\Pi_k(\x_0,t+dt))/\pi_k(\x_0)
=-\frac{1}{\pi_k[\x_0]}\frac{\partial \Pi_k(\x_0,t]}{\partial t}dt,
\end{align*}
since $\pi_k=\P[\Omega_k]$.
That is,
\begin{equation}
\label{fk}
f_k(\x_0,t)=\frac{J_k(\x_0,t)}{\pi_k(\x_0)}.
\end{equation} 
It follows that the Laplace transform of $f_k(\x_0,t)$ is the generator of the moments of the conditional FPT density:
\begin{align}
\label{fkLT}
\E[\e^{-s\calT_k}|1_{\Omega_k}]=\widetilde{f}_k(\x_0,s)=\frac{\widetilde{J}_k(\x_0,s)}{\widetilde{J}_k(\x_0,0)},
\end{align}
and
\begin{equation}
T_k^{(n)}=\E[\calT_k^n|1_{\Omega_k}]=\left . \left (-\frac{d}{ds}\right )^n\E[\e^{-s\calT_k}|1_{\Omega_k}]\right |_{s=0}=\left . \left (-\frac{d}{ds}\right )^n\widetilde{f}_k(\x_0,s)\right |_{s=0}.
\end{equation}
In particular, using equations (\ref{Pik}) and (\ref{fkLT}), the first and second moments $T_k=T_k^{(1)}$ and $T_k^{(2)}$ are
\begin{equation}
\label{mfpt}
\pi_k(\x_0)T_k(\x_0)=-\left .\frac{d\widetilde{f}_k(\x_0,s)}{ds}\right |_{s=0}=\widetilde{\Pi}_k(\x_0,0),
\end{equation}
and
\begin{equation}
\label{fpt2}
\pi_k(\x_0)T_k^{(2)}(\x_0)=\left .\frac{d^2\widetilde{f}_k(\x_0,s)}{ds^2}\right |_{s=0}=-2\left . \frac{d\widetilde{\Pi}_k(\x_0,0)}{ds}\right |_{s=0}.
\end{equation}

\section{Matched asymptotics}

It follows from the above analysis that one way to calculate the splitting probabilities and the Laplace transformed conditional FPT densities (\ref{fkLT}) is to solve Eq. (\ref{master}) in Laplace space. The latter takes the form
\begin{subequations} 
\label{masterLT}
\begin{align}
	D\nabla^2 \p(\x,s|\x_0) -s\p(\x,s|\x_0)&= -\delta(\x-\x_0) , \ \x\in \calU\backslash \calU_a,\\
	 \nabla \p \cdot \n=0, \ \x\in \partial \calU,\  \p(\x,s|\x_0)&= 0,\ \x\in \partial\calU_i .
	\end{align}
\end{subequations}
Equations (\ref{masterLT}) define a boundary value problem that can be analyzed along analogous lines to previous studies of diffusion in 2D domains with small targets \cite{Bressloff08,Coombs09,Ward15,Lindsay16}. We proceed by matching appropriate
`inner' and `outer' asymptotic expansions in the
limit of small target size $\varepsilon\to 0$. Introducing the Green's function of the modified Helmholtz equation according to
\begin{subequations}
\label{GMH}
\begin{align}
	D\nabla^2 G(\x,s|\x_0) -sG(\x,s|\x_0) &=-\delta(\x-\x_0) , \ \x\in \calU,\\
	 \nabla G(\x,s|x_0)\cdot \n  &=0,\ \x \in \partial \calU, \\
	 G(\x,s|\x_0)&=-\frac{1}{2\pi D}\ln|\x-\x_0|+R(\x,s|\x_0),
	\end{align}
	\end{subequations}
where $R$ is the regular part of $G$, we set
\begin{equation}
\p(\x,s|\x_0)=G(\x,s|\x_0)+D^{-1}\q(\x,s|\x_0),\ \x \in \calU\backslash \calU_a,
\end{equation}
with
\begin{subequations} 
\label{masterLT2}
\begin{align}
	D\nabla^2 \q(\x,s|\x_0) -s\q(\x,s|\x_0)&= 0 , \ \x\in \calU\backslash \calU_a,\\
	 \nabla \q \cdot \n=0, \ \x\in \partial \calU,\  \q(\x,s|\x_0)&=\Phi_i:= -G(\x_i,s|x_0)D,\ \x\in \partial\calU_i.
	\end{align}
\end{subequations}
Since we will be carrying out an asymptotic expansion with respect to the small parameter $\nu=-1/{\ln \epsilon}$,
while neglecting all terms of $O(\epsilon)$, we have taken $G(\x,s|\x_0)\approx G(\x_i,s|\x_0)$ on the circle $\partial \Omega_i$.

\begin{figure}[t!]
  \centering
  \includegraphics[width=10cm]{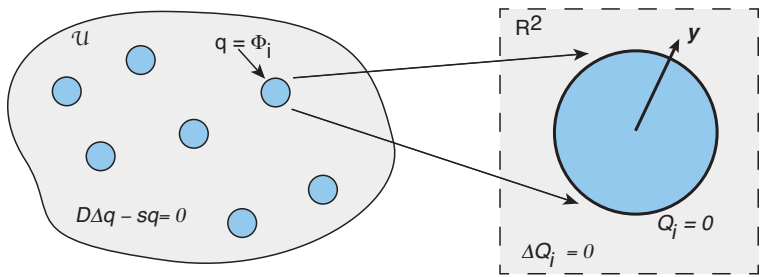}
  \caption{Construction of the inner solution in terms of stretched coordinates $\y=\epsilon^{-1}(\x-{\x}_i)$, where ${\x}_i$ is the center of the $i$-th trap. Rescaled radius is $|\y|=1$ and the region outside the trap is taken to be $\R^2$ rather than the bounded domain $\calU$.}
  \label{fig2}
\end{figure}

\begin{figure}[b!]
  \centering
  \includegraphics[width=11cm]{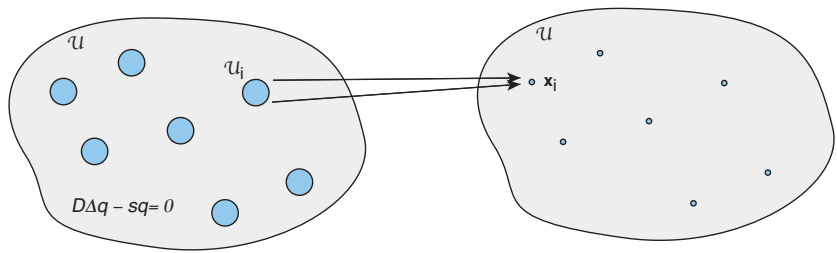}
  \caption{Construction of the outer solution. Each target is shrunk to a single point. The outer solution can be expressed in terms of the Neumann Green's function of the modified Helmholtz equation and then matched with the inner solution around each target.}
  \label{fig3}
\end{figure}

In the inner region around the $i$-th target, we introduce the stretched coordinates $\y=(\x-\x_j)/\epsilon$ and expand the inner solution as (see Fig. \ref{fig2})
\begin{subequations}
\begin{equation}
\q(\x_i+\epsilon \y,s|\x_0)=\nu A_i(\nu,s)Q_i(\y,s|\x_0)+\Phi_i,
\end{equation}
with 
\begin{align}
  \nabla^2_{\y}Q_i &= 0\,, \quad  |\y| > 1 ,\quad Q_i=0 ,\quad |\y |=1 ,
\end{align}
and the far-field condition
\begin{equation}
Q_i\sim \ln|\y|+O(|\y|^{-1}).
\end{equation}
\end{subequations}
(Note that we have dropped the explicit dependence of the unknown coefficients $A_i(\nu,s)$ on $\x_0$ for ease of notation.)
The outer solution is constructed by shrinking each target to a single point (see Fig. \ref{fig3})  and imposing a corresponding singularity condition. The leading order contribution to the outer solution thus satisfies
\begin{subequations} 
\label{outer}
\begin{align}
	D\nabla^2 \q(\x,s|\x_0) -s\q(\x,s|\x_0)&= 0 , \ \x\in \calU\backslash \{\x_1,\ldots,\x_N\},\quad
	 \nabla \q \cdot \n=0, \ \x\in \partial \calU,	\end{align}
	 with
	 \begin{equation}
	 \q(\x,s|\x_0)\sim \Phi_j+\nu A_j(\nu,s)\ln|\x-\x_j|/\epsilon \quad \mbox{as} \ \x\rightarrow \x_j.
	 \end{equation}
\end{subequations}
In other words, we have the following equation for $\q$ on $\calU$:
 \begin{align}
	D\nabla^2 \q(\x,s|\x_0) -s\q(\x,s|\x_0)&= 2\pi \nu D \sum_{j=1}^N A_j(\nu,s)\delta(\x-\x_j),\ \x \in \calU ,\nonumber  \\
	 \nabla \q \cdot \n&=0, \ \x\in \partial \calU.	\end{align}
Applying the divergence theorem to this equation implies that
\begin{equation}
\label{cond}
2\pi \nu D\sum_{j=1}^NA_j(\nu,s)=-s\int_{\calU}\q(\x,s|\x_0)d\x.
\end{equation}
The outer equation then has the solution
\begin{equation}
\q(\x,s|\x_0)=-2\pi \nu D \sum_{j=1}^N A_j(\nu,s)G(\x,s|\x_j).
\end{equation}

We have $N$ unknown coefficients $A_j(\nu,s)$, which are obtained by solving $N$ constraints. The latter are constructed by matching the far-field behavior of the inner solutions $Q_j$ with the near-field behavior of the outer solution $\q$ in a neighborhood of $\calU_j$ for $j=1,\ldots,N$:
\begin{equation}
\label{match1}
2\pi \nu D\sum_{i\neq j}  A_{i}(\nu,s)G(\x_i,s|\x_j)+A_{j}(\nu,s)+2\pi \nu D A_{j}(\nu,s)R(\x_j,s|\x_j)
= -\Phi_j .
\end{equation}
In the case of a single target ($N=1$) this can be solved explicitly as
\begin{equation}
\label{single}
A_1(\nu,s)=\frac{DG(\x_1,s|\x_0)}{1+2\pi \nu DR(\x_1,s|\x_1)}.
\end{equation}
For $N>2$, (\ref{match1}) is a matrix equation that has the formal solution
\begin{equation}
\label{matrix}
A_{i}=-\sum_{j=1}^N[\delta_{i,j}+2\pi \nu {\mathcal G}_{ij}(s)]^{-1}\Phi_j,\quad i=1,\ldots,N,
\end{equation}
where
\[{\mathcal G}_{ij}(s)= G(\x_i,s|\x_j)D,\ \mbox{for } i\neq j;\quad {\mathcal G}_{ii}(s)= R(\x_i,s|\x_i)D.
\]
It is important to note that the non-perturbative solutions (\ref{single}) and (\ref{matrix}) for the coefficients $A_i(\nu)$ effectively sum over all logarithmic terms, which is equivalent to calculating the asymptotic solution for all terms of $O(\nu^k)$ for any $k$. This type of summation was originally obtained by Ward and Keller \cite{Ward93}, and is a common feature of strongly localized perturbations in 2D domains. Since $\nu \rightarrow 0$ more slowly than $\epsilon\rightarrow 0$, the summation over logarithmic terms yields $O(1)$ accuracy with respect to $\epsilon$. (It is also possible to include $O(\epsilon)$ and higher-order terms as illustrated in Ref. \cite{Lindsay15}.)

Finally, substituting the inner solution into the Laplace transform of the flux equation (\ref{J}) and using equation (\ref{fkLT}), we obtain the asymptotic expansion of the conditional FPT densities in Laplace space:
\begin{align}
\pi_k(\x_0)\widetilde{f}_k(\x_0,s)&=\widetilde{J}_k(\x_0,s)=-D \int_{\partial \calU_k} \nabla \widetilde{p}(\x,s|\x_0)\cdot \n d\sigma\nonumber \\
&=-D \int_{\partial \calU_k}\nabla (G(\x,s|\x_0)+D^{-1}\q(\x,s|\x_0))\cdot \n d\sigma\nonumber \\
&= -\int_{\partial \calU_k}\left [D\nabla G(\x,s|\x_0)\cdot \n d\sigma-2\pi \nu A_k(\nu,s)\right ].
\end{align}
Using the fact that
	\[	\int_{\partial \calU_k}\nabla G(\x,s|\x_0)\cdot \n d\sigma=\int_{\partial \calU_k}\nabla G(\x_k,s|\x_0)\cdot \n d\sigma+O(\epsilon),\]
	and noting that $\nabla G(\x_k,s|\x_0)$ is a constant vector, we see that the integral term on the right-hand side vanishes. We thus have our main result, namely, that to leading order in $\epsilon$, 
\begin{equation}
\label{asymfk}
\pi_k(\x_0)\widetilde{f}_k(\x_0,s)=2\pi\nu A_k(\nu,s),
\end{equation}
with $A_k$ given by the solution of equation (\ref{matrix}). Summing both sides of equation (\ref{asymfk}) with respect to $k$ recovers a previous result for the unconditional FPT density \cite{Lindsay16}
\begin{equation}
\sum_{k=1}^N \pi_k(\x_0)\widetilde{f}_k(\x_0,s)=2\pi\nu \sum_{k=1}^N A_k(\nu,s).
\end{equation}

One particular advantage of working with the Laplace transformed FPT densities is that asymptotic expansions of splitting probabilities and FPT moments can be obtained directly by performing a Taylor expansion in the Laplace variable $s$ for fixed $\nu$, rather than separately solving a boundary value problem for each statistical quantity as in previous studies  \cite{Coombs15,Ward15,Lindsay16,Bressloff20}.

\subsection{Splitting probabilities}
Taking the limit $s\rightarrow 0$ in equation (\ref{asymfk}) and using $\widetilde{f}_k(\x_0,0)=1$, yields
\begin{align}
\label{split2}
\pi_k &=2\pi \nu \lim_{s\rightarrow 0}A_k(\nu,s).\end{align}
In order to determine the small-$s$ behavior of $A_k(\nu,s)$, we use the result that
\begin{equation}
\label{Gs}
G(\x,s|\x_0)=\frac{1}{s|\calU|}+G_0(\x,\x_0)+sG_1(\x,\x_0)+o(s),
\end{equation}
where $G_0$ is the modified Neumann Green's function of the diffusion equation:
\begin{subequations}
\label{GM}
\begin{align}
	D\nabla^2 G_0(\x,\x_0)  &=\frac{1}{|\calU|}-\delta(\x-\x_0) , \ \x\in \calU,\\
	 \nabla G_0(\x,\x_0)\cdot \n  &=0,\ \x \in \partial \calU,\quad \int_{\calU}G_0(\x,\x_0)d\x=0, \\
	 G_0(\x,\x_0)&=-\frac{1}{2\pi}\ln|\x-\x_0|+R_0(\x,\x_0),
	\end{align}
	\end{subequations}
	and
	\begin{subequations}
\label{GM1}
\begin{align}
	D\nabla^2 G_1(\x,\x_0)  -G_0(\x,\x_0)&=0 , \ \x\in \calU,\\
	 \nabla G_1(\x,\x_0)\cdot \n  &=0,\ \x \in \partial \calU,\quad \int_{\calU}G_1(\x,\x_0)d\x=0.	\end{align}
	\end{subequations}
	Note that the solution of equation (\ref{GM1}) is non-singular and takes the form
	\begin{equation}
	G_1(\x,\x_0)=-\int_{\calU}G_0(\y,\x)G_0(\y,\x_0)d\y.
	\end{equation}
Taking the limit $s\rightarrow 0$ in equation (\ref{cond}) yields
\begin{align}
\label{cond2}
2\pi \nu \lim_{s\rightarrow 0}\sum_{j=1}^NA_j(\nu,s)&=-D^{-1}\lim_{s\rightarrow 0} s\int_{\calU}\q(\x,s|\x_0)d\x\nonumber \\
&=-\lim_{s\rightarrow 0} s\int_{\calU}[\p(\x,s|\x_0)-G(\x,s|\x_0)]d\x\nonumber \\
&=-\lim_{t\rightarrow \infty}\int_{\calU}[p(\x,t|\x_0)-G(\x,t|\x_0)]d\x=1.
\end{align}
Hence, the coefficient $A_k(\nu,s)$ has the small-$s$ expansion
\begin{equation}
\label{anu}
A_k(\nu,s)=\frac{1}{2\pi \nu N}+A_k(\nu)+s\chi_k(\nu)+s^2 \zeta_k(\nu)+O(s^3)
\end{equation}
for some $\chi_k(\nu),\zeta_k(\nu)$, with the constraint
\begin{equation}
\sum_{k=1}^NA_k(\nu)=0.
\end{equation}
It immediately follows that the splitting probability is
\begin{equation}
\label{split3}
\pi_k=\frac{1}{N}+2\pi \nu A_k(\nu).
\end{equation}

It remains to determine the coefficients $A_k(\nu)$ by substituting equations (\ref{Gs}) and (\ref{anu}) into the matching equation (\ref{match1}) and taking the limit $s\rightarrow 0$. This yields
\begin{align}
&\sum_{j\neq k} \left [2\pi \nu A_{j}(\nu)+\frac{1}{N}\right ]G_0(\x_j,\x_k)+\frac{1}{D}\left (A_{k}(\nu)+\frac{1}{2\pi \nu N}\right )
\nonumber\\
&\qquad +\left [2\pi \nu A_{k}(\nu)+\frac{1}{N}\right ]R_0(\x_k,\x_k) =G_0(\x_k,\x_0)-\frac{2\pi \nu}{|\calU|}\chi(\nu),
\label{match2}
\end{align}
where $\chi(\nu)=\sum_{k=1}^N\chi_k(\nu)$.
Summing both sides of equation (\ref{match2}) with respect to $k$ and using $\sum_kA_k(\nu)=0$ gives
\begin{align}
\label{chi}
-\frac{2\pi \nu }{|\calU|}\chi(\nu)&=\frac{1}{2\pi \nu D N}+\frac{1}{N}\sum_{i,j=1}^N\left [2\pi \nu A_{i}(\nu)+\frac{1}{N}\right ]{\mathcal G}_{ij}^{(0)}-\frac{1}{N}\sum_{j=1}^NG_0(\x_j,\x_0),
\end{align}
 with
\[{\mathcal G}_{ij}^{(0)}= G_0(\x_i,\x_j),\ \mbox{for } i\neq j;\quad {\mathcal G}_{ii}^{(0)}= R_0(\x_i,\x_i).
\]
We can thus eliminate $\chi(\nu)$ from equation (\ref{match2}) such that 
\begin{align}
\label{match3}
&\sum_{j=1}^N\left [2\pi \nu A_{j}(\nu)+\frac{1}{N}\right ]{\mathcal G}_{jk}^{(0)}-\frac{1}{N}\sum_{j,l=1}^N\left [2\pi \nu A_{j}(\nu)+\frac{1}{N}\right ]{\mathcal G}_{jl}^{(0)}+\frac{A_k(\nu)}{D}
\nonumber\\
&\qquad =G_0(\x_k,\x_0)-\frac{1}{N}\sum_{j=1}^NG_0(\x_j,\x_0).
\end{align}
Equations (\ref{split3}) and (\ref{match3}) determine the asymptotic expansion of the splitting probabilities.

\subsection{First passage times}

An analogous asymptotic analysis can be used to determine the moments of the conditional FPT density. 
These are given by
\begin{equation}
T_k^{(n)}=\left . \left (-\frac{d}{ds}\right )^n\widetilde{f}_k(\x_0,s)\right |_{s=0}.
\end{equation}
For example, equations (\ref{Pik}) and (\ref{mfpt}) imply that
\begin{align}
 \pi_k(\x_0)T_k(\x_0)&=\widetilde{\Pi}_k(\x_0,0)=\lim_{s\rightarrow 0}\frac{\pi_k(\x_0)-\widetilde{J}_k(\x_0,s)}{s}=\lim_{s\rightarrow 0}\frac{\widetilde{J}_k(\x_0,0)-\widetilde{J}_k(\x_0,s)}{s}\nonumber \\
 &=-2\pi \nu \lim_{s\rightarrow 0}\frac{dA_k(\nu,s)}{ds}= -2\pi \nu \chi_k(\nu).
 \label{pea}
\end{align}
Similarly, equations (\ref{Pik}) and (\ref{fpt2}) give
\begin{align}
\frac{1}{2}\pi_k(\x_0)T_k^{(2)}(\x_0)=-\left . \frac{d\widetilde{\Pi}_k(\x_0,0)}{ds}\right |_{s=0}=-\lim_{s\rightarrow 0}\frac{d}{ds} \frac{\widetilde{J}_k(\x_0,0)-\widetilde{J}_k(\x_0,s)}{s}=2\pi \nu \zeta_k(\nu).
\end{align}
The coefficients $\chi_k(\nu)$ and $\zeta_k(\nu)$ can be determined by expanding the matching condition (\ref{match1}) to $O(s^2)$. Here we simply state the results for a single target:
\begin{equation}
\label{single2}
2\pi \nu \chi_1(\nu)=|\calU|\left [G_0(\x_1,\x_0)-R_0(\x_1,\x_1)-\frac{1}{2\pi \nu D}\right ],
\end{equation}
and
\begin{equation}
\label{single3}
2\pi \nu \zeta_1(\nu)=|\calU|\left [G_1(\x_1,\x_0)-G_1(\x_1,\x_1)-\frac{\chi_1(\nu)[1+2\pi \nu DR_0(\x_1,\x_1)]}{D}\right ].
\end{equation}

\section{Stochastic resetting}

As our first application of the result (\ref{asymfk}) to extended narrow capture problems, suppose that we introduce a stochastic resetting protocol into the search process given by equation (\ref{master}). That is, prior to being absorbed by one of the targets, the particle can reset to a fixed location $\x_r$ at a random sequence of times generated by an exponential probability density $\psi(\tau)=r\e^{-r\tau}$, where $r$ is the resetting rate. The probability that no resetting has occurred up to time $\tau$ is then $\Psi(\tau)=1-\int_0^{\tau}\psi(s)ds=\e^{-r\tau}$. In the following we identify $\x_r$ with the initial position by setting $\x_0=\x_r$, as illustrated in Fig. \ref{fig1} for reset before capture. For simplicity, we will assume that the resetting is instantaneous and there are no refractory delays. (Extensions to finite return times and other delays can be found elsewhere \cite{Pal19,Pal20,Bressloff20m}.) Using renewal theory it can be shown that the splitting probabilities and conditional FPT densities with resetting, which are distinguished by the subscript $r$, can be expressed in terms of statistical quantities without resetting as follows \cite{Bressloff20}:
\begin{align}
\pi_{r,k}(\x_r)&=\frac{\pi_{k}(\x_0)-r\widetilde{\Pi}_{k}(\x_0,r)}{ 
1-r\widetilde{Q}(\x_0,r)}=\frac{\widetilde{J}_{k}(\x_0,r)}{\sum_{j=1}^N\widetilde{J}_{j}(\x_0,r)}=\frac{A_k(\nu,r)}{\sum_{j=1}^NA_j(\nu,r)},
\label{Piee}
\end{align}
and
\begin{align}
\label{fcond}
\pi_{r,k}(\x_0)\widetilde{f}_{r,k}(\x_0,s)&=\frac{\pi_{k}(\x_0)-(r+s)\widetilde{\Pi}_{k}(\x_0,r+s)}{1-r\widetilde{Q}(\x_0,r+s)}=\frac{2\pi \nu (r+s)A_k(\nu,r+s)}{s+2\pi \nu r\sum_{j=1}^NA_j(\nu,r+s)}.
\end{align}
We have used equations (\ref{Pik}) and (\ref{asymfk}). Moreover, differentiating equation (\ref{fcond}) with respect to $s$ shows that the unconditional MFPT with resetting is
\begin{equation}
\label{Tcond}
T_{r}(\x_0)=\frac{\widetilde{Q}(\x_0,r)}{1-r\widetilde{Q}(\x_0,r)}=\frac{1}{r}\frac{1-2\pi \nu \sum_{j=1}^NA_j(\nu,r)}{2\pi \nu \sum_{j=1}^NA_j(\nu,r)}.
\end{equation}
In the case of a single target, we have $\pi_{r,1}(\x_0)=1$ and equation (\ref{single}) implies that
\begin{equation}
\label{Tsingle}
T_{r}(\x_0)=\frac{1}{r}\frac{1-2\pi \nu  A_1(\nu,r)}{2\pi \nu A_1(\nu,r)}=\frac{1}{r}\frac{1+2\pi \nu D[R(\x_1,s|\x_1)-G(\x_1,s|\x_0)]}{2\pi \nu DG(\x_1,s|\x_0)}.
\end{equation}

Let us first consider the small-$r$ regime, which was previously analyzed in \cite{Bressloff20}. Taylor expanding equation (\ref{Piee}) in $r$ for fixed $\nu$, we have 
\begin{align}
\pi_{r,k}(\x_r)&=\frac{\pi_k(\x_0)+rA_k'(\nu,0)+o(r)}{1+r\sum_{j=1}^NA_j'(\nu,0)+o(r)}
=\frac{\pi_k(\x_0)[1-rT_k(\x_0)]+o(r)}{1-rT(\x_0)+o(r)}\nonumber \\
&=\pi_k(\x_0)+r\pi_k(\x_0)(T(\x_0)-T_k(\x_0))+o(r),
\label{Piee2}
\end{align}
where $A_k'$ indicates differentiation with respect to the Laplace variable. 
Hence, we recover the condition that to $O(r)$, resetting at a rate $r$ will increase the splitting probability to the $k$-th target provided that $T(\x_0)>T_k(\x_0)$.
Similarly, Taylor expanding equation (\ref{Tcond}) gives
\begin{align}
\label{Tcond2}
T_{r}(\x_0)&=-\frac{2\pi \nu \sum_{k=1}^N\left [A_k'(\nu,0)+rA_k''(\nu,0)/2+o(r)\right ]}{1+2\pi \nu r\sum_{j=1}^NA_j'(\nu,0)+o(r)}
\\
&=\frac{T(\x_0)-rT^{(2)}(\x_0)/2+o(r)}{1-rT(\x_0)+o(r)}\nonumber \\
&=T(\x_0)+r[T(\x_0)^2-T^{(2)}(\x_0)/2]+o(r).\nonumber
\end{align}
where $T^{(2)}(\x_0)$ is the unconditional second moment without resetting. We see that the unconditional MFPT will be reduced in the small-$r$ regime provided that the coefficient of variation (CV) satisfies the inequality $\sigma_T(\x_0)/T(\x_0)>1$, where $\sigma_T=\sqrt{T^{(2)}-T^2}$. 

In the case of a single target, we can derive an explicit expression for the above inequality using equations (\ref{single2}) and (\ref{single3}). That is,
\begin{align}
&T(\x_0)^2-T^{(2)}(\x_0)/2=|\calU|^2\left [G_0(\x_1,\x_0)-R_0(\x_1,\x_1)-\frac{1}{2\pi \nu D}\right ]^2\nonumber \\
&\quad -\frac{|\calU|}{2}\left [G_1(\x_1,\x_0)-G_1(\x_1,\x_1)-\frac{\chi_1(\nu)[1+2\pi \nu DR_0(\x_1,\x_1)]}{D}\right ]\nonumber \\
&=\frac{|\calU|^2}{(2\pi \nu D)^2}-\frac{|\calU|^2}{\pi \nu D}[G_0(\x_1,\x_0)-R_0(\x_1,\x_1)]\nonumber \\
&\quad +\frac{|\calU|^2}{2\pi \nu D}\left [G_0(\x_1,\x_0)-R_0(\x_1,\x_1)-\frac{1}{2\pi \nu D}\right ] \bigg [1+2\pi \nu DR_0(\x_1,\x_1)\bigg ]  +O(1)\nonumber\\
&=-\frac{|\calU|^2}{\pi \nu D}G_0(\x_1,\x_0)+O(1).\end{align}
Hence, to leading order, the MFPT will be reduced in the small-$r$ regime provided that $G_0(\x_1,\x_0)>0$. The generalization to the multiple target case is
 $\sum_{j=1}^NG_0(\x_j,\x_0)>0$. Note that in \cite{Bressloff20} this result was derived by separately solving the BVPs for $T$ and $T^{(2)}$.

Another major advantage of working with the full FPT densities is that the expressions (\ref{Piee}) and (\ref{Tcond}) for the splitting probabilities and unconditional MFPT hold for all resetting rates $r$. Hence, the calculation of the splitting probabilities and FPTs outside the small-$r$ regime reduces to the problem of obtaining accurate numerical or analytical approximations of the modified Helmholtz Green's function $G(\x,r|\x')$. This particular issue has been addressed by Lindsay {\em et al.}  \cite{Lindsay16}, and we adapt their results to the current problem. It turns out that an important step in the evaluation of the Green's function is to decompose $G$ as the sum of the free-space Green's function and a regular boundary-dependent part:
\begin{equation}
G(\x,s|\x')=\frac{1}{2\pi D}K_0(\sqrt{s/D}|\x-\x'|)+\widehat{R}(\x,s|\x'),
\end{equation}
where $K_0(z)$ is the modified Bessel function of the second kind and $\widehat{R}$ is nonsingular as $\x \rightarrow \x'$. Using a boundary integral method, it can be shown that for $|\x-\x_0|=O(1)$ and large $\sqrt{s/D}$, the boundary contributions to $\widehat{R}$ are exponentially small so that \cite{Lindsay16}
\begin{align}
G(\x,s|\x_0)&\sim \frac{1}{2\pi D}K_0(\sqrt{s/D}|\x-\x_0|), \ \x\neq \x_0,\nonumber \\
\widehat{R}(\x_0,s|\x_0)&\sim -\frac{1}{2\pi D}(\ln \sqrt{s/D}-\ln 2+\gamma_c),
\end{align}
where $\gamma_c\approx 0.5772$ is Euler's gamma constant. 
We now observe that the off-diagonal matrix terms in (\ref{match1}) are exponentially smaller than the diagonal elements, so that to leading order,
\begin{equation}
\label{Ak}
A_k(\nu,s)=\frac{(2\pi)^{-1} K_0(\sqrt{s/D}|\x_k-\x_0|)}{1-\nu(\ln \sqrt{s/D}-\ln 2+\gamma_c)},\quad \sqrt{s/D} \gg 1 .
\end{equation}
Hence, under the boundary-free approximation, $A_k(\nu,r)$ depends on the distances of the targets from $\x_0$ but is independent of the shape of the domain and the absolute locations of the targets. Numerically it has been shown that such an approximation remains valid even at intermediate values of $r$ (or equivalently at intermediate times) provided that $\x_0$ and $\x_j$, $j=1,\dots,N$, are not close to the boundary and there are no bottlenecks separating the targets from $\x_0$. This result is reinforced in the presence of resetting.

\begin{figure}[t!]
\begin{center} 
\includegraphics[width=12cm]{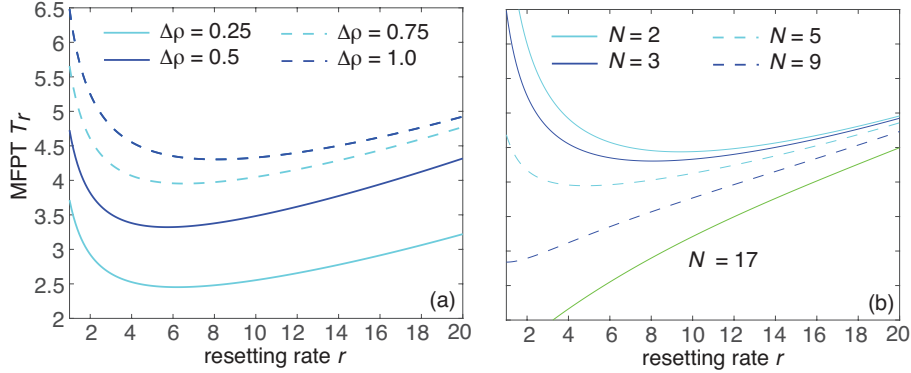} 
\caption{Plot of unconditional MFPT $T_r(\x_0)$ for $\sqrt{r/D} >1$ and $N$ targets  with $|\x_n-\x_0|=\rho_0+(n-1)\Delta \rho$. (a) $\rho_0=0.5$, $N=3$ and various $\Delta \rho$. (b) $\rho_0=0.5$, $\Delta \rho=1$ and various $N$. Other parameters are $D=1$ and $\nu=0.1$. }
\label{fig4}
\end{center}
\end{figure}

Substituting for $A_k$ into equations (\ref{Piee}) and (\ref{Tcond}) yields the finite-$r$ approximations
\begin{align}
\pi_{r,k}(\x_0)&\sim \frac{K_0(\sqrt{r/D}|\x_k-\x_0|)}{\sum_{j=1}^NK_0(\sqrt{r/D}|\x_j-\x_0|)},
\label{Piee3}
\end{align}
and
\begin{align}
\label{Tcond3}
T_{r}(\x_0)&\sim \frac{1}{r}\frac{1 -\nu(\ln \sqrt{r/D}-\ln 2+\gamma_c) -\nu \sum_{j=1}^NK_0(\sqrt{r/D}|\x_j-\x_0|)}{  \nu \sum_{j=1}^NK_0(\sqrt{r/D}|\x_j-\x_0|)} .
\end{align}
 We can then use equation (\ref{Tcond3}) to explore how the unconditional MFPT varies with the resetting rate $r$, the number of targets and the distribution of distances $|\x_i-\x_0|$, $i=1,\ldots,N$. In Fig. \ref{fig4} we plot the finite-$r$ expression for $T_r$ as a function of $r$ for a set of $N$ targets whose distances from $\x_0$ are of the form $\rho_n:=|\x_b-\x_0| =\rho_0+(n-1)\Delta \rho$. Since the approximations (\ref{Piee3}) and (\ref{Tcond3}) break down for small $r$ we impose the restriction $r/D >1$. For a range of parameter values we find that $T_r$ has a minimum at an optimal resetting rate $r_{\rm opt}$ that depends on $N$, $\Delta \rho$ and $\rho_0$. In all cases $T_r\rightarrow \infty$ as $r\rightarrow \infty$, which reflects the fact that if the particle resets too often then it never has the chance to reach even the closest target. The divergence of $T_r$ can be explored further using the asymptotic expansion
\begin{subequations}
\begin{align}
K_0(\  z)&\sim \sqrt{\frac{\pi}{2  z}}\e^{-z}\left [1-\frac{1}{8  z}+O( z^{-2}).\right ],\quad z\rightarrow \infty,
\end{align}
\end{subequations}
which implies that
\begin{align}
\pi_{r,k}(\x_0)&\sim \frac{\e^{-\sqrt{r/D}|\x_k-\x_0|}/\sqrt{|\x_k-\x_0|}}{\sum_{j=1}^N\e^{-\sqrt{r/D}|\x_j-\x_0|}/\sqrt{|\x_j-\x_0|}},\ \sqrt{r/D} \gg 1 ,
\label{Piee5}
\end{align}
and
\begin{align}
\label{Tcond5}
T_{r}(\x_0)
&\sim \frac{1}{r}\frac{\sqrt{2/\pi}- \nu \sum_{j=1}^N\e^{-\sqrt{r/D}|\x_j-\x_0|}/\sqrt{\sqrt{r/D}|\x_j-\x_0|}}{ \nu \sum_{j=1}^N\e^{-\sqrt{r/D}|\x_j-\x_0|}/\sqrt{\sqrt{r/D}|\x_j-\x_0|}} ,\ \sqrt{r/D} \gg 1.
\nonumber 
\end{align}
If we now order the targets such that
\[|\x_1-\x_0| <|x_2-\x_0|\leq \ldots \leq |\x_N-\x_0|,\]
then the target closest to the reset point $\x_0$ dominates in the large-$r$ regime and
\begin{equation}
\label{eqT}
T_r(\x))\sim \frac{\sqrt{\sqrt{r /D}|\x_1-\x_0}|}{\nu r}\sqrt{\frac{2}{\pi}} \e^{\sqrt{r/D}|\x_1-\x_0|},\quad r\rightarrow \infty.
\end{equation}
Equation (\ref{eqT}) establishes that the unconditional MFPT diverges exponentially.

\section{Accumulation of target resources}

Turning to our second example of an extended narrow escape problem, suppose that, rather than being permanently absorbed or captured by a target, the particle delivers a discrete packet of some resource (burst event) to the target and then returns to $\x_0$, initiating another round of search-and-capture. The return to $\x_0$ is distinct from resetting before capture considered in section 4, since the latter can occur at any time, see Fig. \ref{fig1}. (One could also consider a combination of both forms of resetting \cite{Bressloff20}.) The sequence of bursts resulting from multiple rounds of search-and-capture leads to an accumulation of packets within the targets, which we assume is counteracted by resource degradation at some rate $\gamma$. 
Furthermore, suppose that the total time for the particle to unload its cargo, return to $\x_0$ and start a new search process is given by the random variable $ {\tau}$ with waiting time density $\rho( {\tau})$, which for simplicity is taken to be independent of the location of the targets (under the assumption of instantaneous reset).
Let $M_k(t)$ be the number of resource packets in the $k$-th target that have not yet degraded at time $t$. As we have shown elsewhere \cite{Bressloff19,Bressloff20a,Bressloff20}, the accumulation of resources within the targets can be analyzed by reformulating the multiple search-and-capture model as a G/M/$\infty$ queuing process \cite{Takacs62,Liu90}. Here we simply quote the results for the steady-state mean and variance of $M_k$. 

First, the mean is given by
 \begin{equation} \label{mean}
\overline{M}_k(\x_0,\gamma)  =\frac{\pi_k(\x_0)}{  \gamma[T(\x_0)  +\tau_{\rm cap}]},
\end{equation}
where $\pi(\x_0)$ and $  T(\x_0)$ are the splitting probabilities and unconditional MFPT of a single search-and-capture event, and $\tau_{\rm cap}=\int_0^{\infty}\tau\rho(\tau)d\tau$ is the mean loading/unloading time. Hence, the relative distribution
of resources across the population of competing targets is determined by the splitting probabilities $\pi_k(\x_0)$, whereas the total number of resources depends on the unconditional MFPT $T(\x_0)$, that is, 
\[\overline{M}_{\rm tot}(\x_0,\gamma) =\sum_{k=1}^N\overline{M}_k(\x_0 ,\gamma)=\frac{1}{\gamma(T(\x_0)+\tau_{\rm cap})}.\]
On the other hand, the variance (and higher-order moments of the resource distribution) depend on the full conditional FPT densities in Laplace space that were calculated in section 3. In particular, the variance takes the form
\begin{align}
\label{var}
 \mbox{Var}[M_k(\x_0,\gamma)]=\overline{M}_k(\x_0,\gamma)  \left [ \frac{\pi_{k}(\x_0)\widetilde{\rho}(\gamma)\widetilde{f}_{k}(\x_0,\gamma)}{1-\sum_{j=1}^N\pi_{j}(\x_0 )\widetilde{\rho}(\gamma)\widetilde{f}_{j}(\x_0,\gamma)}+1- \overline{M}_k(\x_0,\gamma) 
\right ],
\end{align}
where $\widetilde{f}_{k}(\x_0,\gamma)$ are the conditional FPT densities.
Using equation (\ref{asymfk}), we see that
\begin{align}
\label{var2}
 \mbox{Var}[M_k(\x_0,\gamma)]=\overline{M}_k(\x_0,\gamma)   \left [ \frac{2\pi \nu \widetilde{\rho}(\gamma) A_k(\nu,\gamma)}{1-2\pi \nu \widetilde{\rho}(\gamma)\sum_{j=1}^N  A_j(\nu,\gamma)}+1-\overline{M}_k(\x_0,\gamma)   
\right ].
\end{align}
It turns out that we can express the variance of the target resource distribution in terms of quantities associated with a search process with stochastic resetting at a rate $\gamma$.
That is, setting $r=\gamma$ in equations (\ref{Piee}) and (\ref{Tcond}), and rearranging, we have
\[\frac{A_k(\nu,\gamma)}{\sum_{j=1}^NA_j(\nu,\gamma)}=\pi_{\gamma,k}(\x_0),\quad 2\pi \nu \sum_{j=1}^N  A_j(\nu,\gamma)=\frac{1}{1+\gamma T_{\gamma}(\x_0)}.\]
Hence, equation (\ref{var2}) can be written in the form
\begin{align}
\label{var3}
 \mbox{Var}[M_{k}(\x_0,\gamma)]
&=\overline{M}_k(\x_0,\gamma)   \left [\frac{  {\pi}_{\gamma,k}(\x_0)\widetilde{\rho}(\gamma)}{\gamma T_{\gamma}(\x_0)+1-\widetilde{\rho}(\gamma)}+1-\overline{M}_k(\x_0,\gamma) \right ] .
\end{align}
In the absence of loading/unloading delays, $\widetilde{\rho}(\gamma)=1$ and $\tau_{\rm cap}=0$, we obtain the simpler expression
\begin{align}
\label{var4}
& \mbox{Var}[M_{k}(\x_0,\gamma)]
 =\frac{\pi_{k}(\x_0)}{\gamma T(\x_0)  } \left [\frac{\pi_{\gamma,k}(\x_0)}{\gamma T_{\gamma}(\x_0)  }-\frac{\pi_{k}(\x_0)}{\gamma T(\x_0)  } +1 \right ] .
\end{align}
Note that $\pi_k(\x_0)=\pi_{0,k}(\x_0)$ and $T(\x_0)=T_0(\x_0)$.
In conclusion, the steady-state variance depends on the splitting probabilities and unconditional MFPTs for a search process without resetting and one with resetting at the degradation rate $\gamma$. 

We now explore the $\gamma$-dependence of the Fano factor (without delays), which we write as
\begin{equation}
\label{fano}
F_k(\x_0,\gamma)=\frac{\mbox{Var}[M_{k}(\x_0,\gamma)]}{\overline{M}_{k}(\x_0,\gamma)}=\frac{2\pi \nu A_k(\nu,\gamma)}{1-2\pi \nu \sum_{j=1}^N  A_j(\nu,\gamma)}-\frac{\pi_{k}(\x_0)}{\gamma T(\x_0)  }  +1 .
\end{equation}
The Fano factor is a natural quantity to consider in the case of a queuing process, since the latter is an example of a counting process that tracks discrete events. The best known example of a counting process is the Poisson process, which is Markovian and has a Fano factor of one. In applications this is often used as a baseline to characterize the noise in a counting process. First, taking the limit $\gamma\rightarrow \infty$ in equation (\ref{fano}) implies
\begin{equation}
\label{cong}
 \lim_{\gamma \rightarrow \infty} F_k(\x_0,\gamma)=1.
\end{equation}
Next, in order to determine $F_k(\x_0,\gamma)$ in the limit $\gamma \rightarrow 0$, we use equation (\ref{anu}): 
\begin{align}
\label{fano2}
F_k(\x_0,\gamma)&=\left [- \frac{1/N+2\pi \nu [A_k(\nu)+\gamma \chi_k(\nu)+o(\gamma)]}{2\pi \nu \sum_{j=1}^N  [A_j(\nu)+\gamma \chi_j(\nu)+\gamma^2\zeta_k(\gamma)+o(\gamma^2)]}-\frac{\pi_{k}(\x_0)}{\gamma T(\x_0)  } +1 \right ]\\
&=\left [\frac{\pi_k(\nu)+2\pi \nu \gamma \chi_k(\nu)+o(\gamma)}{\gamma [T(\x_0)+\gamma T^{(2)}(\x_0)/2+o(\gamma^2)]}-\frac{\pi_{k}(\x_0)}{\gamma T(\x_0)  } +1 \right ]\nonumber \\
&=1-\pi_k(\x_0)\left [\frac{T_k(\x_0)}{T(\x_0)}-\frac{1}{2}\frac{T^{(2)}(\x_0)}{T(\x_0)^2}\right ]+O(\gamma).
\end{align}
Thus, in the limit $\gamma \rightarrow 0$, the Fano factor depends on the splitting probabilities and low-order moments of the FPT densities.

\begin{figure}[b!]
\begin{center} 
\includegraphics[width=12cm]{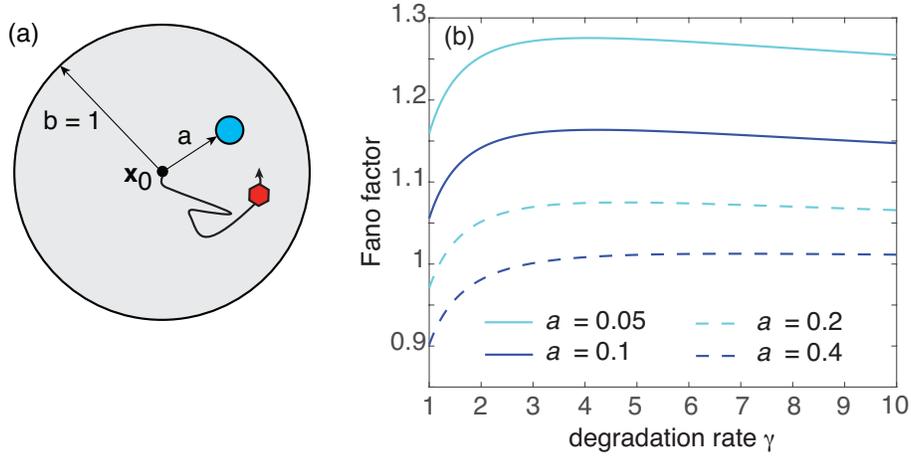} 
\caption{(a) Single target in a circular search domain of unit radius. The target is located at a distance $a$ from the center of the circle. Each time the particle is captured by the target it delivers a resource packet and immediately returns to $\x_0=0$ to initiate another round of search-and-capture. (b) Plot of steady-state Fano factor as a function of the resource degradation rate and various radii $a$. Other parameter values are $D=1$, $\nu=0.1$}
\label{fig5}
\end{center}
\end{figure}

However, we can also determine the Fano factor for finite values of $\gamma$ using the approximations (\ref{Piee3}) and (\ref{Tcond3}):
\begin{align}
F_k(\x_0,\gamma)&=\frac{  \nu K_0(\sqrt{\gamma/D}|\x_k-\x_0|)}{1-\nu(\ln \sqrt{\gamma/D}-\ln 2+\gamma_c)-  \nu \sum_{j=1}^N  K_0(\sqrt{\gamma/D}|\x_j-\x_0|)}\nonumber \\
&\quad-\frac{\pi_{k}(\x_0)}{\gamma T(\x_0)  }  +1. 
\label{fan}
\end{align}
Since $\pi_k(\x_0)$ and $T(\x_0)$ depend on the Green's function $G_0$ of the diffusion equation, which does not decay exponentially to the boundary, we need to specify the shape of the search domain. For the sake of illustration, consider a single target at a radial distance $a$ from the center of a circular search domain with unit radius, see Fig. 5(a), with $a$ sufficiently small so that the approximation (\ref{fan}) holds. For a single target, $\pi_1(\x_0)=1$ and
\begin{equation}
\frac{1}{T(\x_0)}=\frac{2\pi \nu D}{ |\calU|}\frac{1}{1-2\pi \nu D [G_0(\x_1,\x_0)-R_0(\x_1,\x_1)]},
\end{equation}
where $|\calU|=\pi$ and we have used equation (\ref{single2}). For the unit circle with $\x_0=0$,
\begin{equation}
G_0(\x,0)=\frac{1}{2\pi}\left [-\ln |\x|+\frac{|\x|^2}{2}-\frac{3}{4}\right ],\quad R_0(\x,\x)=-\frac{3}{8\pi}.
\end{equation}
Hence, the MFPT is 
\begin{equation}
T=T(a)=\frac{1+ \nu D [\ln a-a^2/2]}{2 \nu D}{1- \nu D [\ln a-a^2/2]},
\end{equation}
and the Fano factor takes the form $F_1=F(a,\gamma)$ with
\begin{align}
F(a,\gamma)&=\frac{  \nu K_0(\sqrt{\gamma/D}a)}{1-\nu(\ln \sqrt{\gamma/D}-\ln 2+\gamma_c)-  \nu   K_0(\sqrt{\gamma/D}a)}-\frac{1}{\gamma T(a)  }  +1. 
\label{fan2}
\end{align}
In Fig. \ref{fig5}(b) we plot $F(a,\gamma)$ as a function of the degradation rate $\gamma$ for various radii $a$. It can be seen that $F\rightarrow 1$ as $\gamma\rightarrow \infty$. Moreover, the Fano factor increases as $a$ decreases, that is, as the distance from the reset point decreases.

\section{Conclusion}

In this paper we used asymptotic PDE methods to analyze extended 2D narrow capture problems. The classical version is based on diffusion of a particle in a bounded domain with one or more small $O(\epsilon)$ absorbing interior targets or traps. First, we developed an efficient method for obtaining asymptotic expansions of statistical quantities such as splitting probabilities and moments of the conditional FPT densities by carrying out a Taylor expansion of the latter in Laplace space. This avoids having to solve a separate BVP for each quantity. We then used our asymptotic analysis of the FPT densities to determine the effects of stochastic resetting on the unconditional MFPT, extending previous results in the small-$r$ regime. We exploited the exponential-like asymptotic decay of the Green's function for the modified Helmholtz equation in order to construct boundary-free approximations of statistical quantities that hold for intermediate and large values of the resetting rate $r$. This allowed us to identify target configurations where the MFPT is minimized at an optimal resetting rate. Finally, we investigated the size of fluctuations in the steady-state distribution of target resources under multiple rounds of search-and-capture. We proceeded by expressing the Fano factor for the number of resources in a target in terms of the splitting probabilities and unconditional MFPTs for a search process with resetting at the degradation rate $\gamma$. In future work we hope to develop further the theory of extended narrow capture problems. One obvious example is to consider three-dimensional search processes. The major difference from 2D is that the associated Green's functions have singularities of the form $1/|\x|$ rather than $-\ln|\x|$ as $\x|\rightarrow 0$. Another example is to consider partially absorbing targets, whereby the sets $\calU_k$ in Fig. \ref{fig1} are treated as partially absorbing domains in which the flux into the $k$-th target is of the form $J_k(\x_0,t)=\kappa\int_{\calU_k}p(\x,t|\x_0)d\x$, where $\kappa$ is an absorption rate. 

\vskip6pt

\enlargethispage{20pt}








\end{document}